\documentstyle[12pt]{article}
\advance\textheight by \topskip
\def\fun#1#2{\lower3.6pt\vbox{\baselineskip0pt\lineskip.9pt
  \ialign{$\mathsurround=0pt#1\hfil##\hfil$\crcr#2\crcr\sim\crcr}}}
\relax
\begin{document}
\begin{flushright}
SU--ITP--96--43\\
astro-ph/9610077\\
October 10, 1996\\
\end{flushright}
\vskip 2 cm
\begin{center}
{\LARGE\bf Prospects of Inflationary Cosmology}\vskip 1.7cm

 {\bf Andrei Linde} \footnote{Round Table Discussion  at the conference
``Critical Dialogues in Cosmology,'' Princeton, May 1996}

\vskip 1.5mm
Department of Physics, Stanford University, \\
Stanford, CA 94305--4060, USA
\end{center}
\vskip 1cm

{\centerline{\large\bf Abstract}}
\begin{quotation}
 In this review I briefly describe the evolution of the inflationary theory
from the scenario based on the idea of supercooling and expansion in the false
vacuum toward the theory of eternally expanding self-reproducing inflationary
universe. I describe   recent development  of inflationary cosmology with
$\Omega \not = 1$, and then discuss some issues related to the possibility to
verify   inflation by comparing its predictions with observational data.  I
argue that it is possible to verify and disprove many particular models of
inflationary cosmology, but it is very difficult   to kill the basic idea of
inflation. It seems that the best (and may be even the only) way to do so is to
suggest a better cosmological theory.
\end{quotation}
 \newpage

\section{Various versions of inflationary cosmology}

One of the highlights of the conference ``Critical Dialogues in Cosmology'' in
Princeton was a discussion of inflationary cosmology by Alan Guth and  Bill
Unruh, see refs. \cite{Alan} and \cite{Bill} in the proceedings. Not
surprisingly, I mainly agree with Alan, even though I must admit that some
critical comments made by Unruh were quite appropriate. However, most of these
comments were addressed to the version of inflationary theory which died almost
15 years ago.  To avoid misunderstandings, one should be more specific
describing inflationary cosmology.

The first semi-realistic inflationary model was proposed by Alexei Starobinsky
in 1979 \cite{Star}. This model was rather complicated, it did not aim on
solving homogeneity, horizon and monopole problems, but it worked. The theory
of density perturbations in this model was first developed by Mukhanov and
Chibisov \cite{Mukh}, and it practically coincided with the corresponding
theory developed a year later in the context of new inflation \cite{Hawk}.

A much simpler model with a very clear physical motivation was proposed by Alan
Guth  in 1981 \cite{Guth}.  His model was so nice  that even now all textbooks
on astronomy and all popular books    describe   inflation as an exponential
expansion of the universe in a supercooled false vacuum state. This is a
seductively simple but incorrect  way to explain the essence of inflation.
Exponential expansion in the false vacuum  in a certain sense is false:  de
Sitter space  with a constant vacuum energy density  equally well can be
considered  as expanding,   contracting, or static, depending on the choice of
a coordinate system. The absence of a preferable hypersurface of decay of the
false vacuum was the main reason of failure of the old inflationary theory.

This problem was resolved in the new inflationary theory \cite{New}. In this
theory, just like in the Starobinsky model,  inflation continues away from the
false vacuum. Importance of this fact should not be overlooked:   density
perturbations produced during inflation   are inversely proportional to $\dot
\phi$, where $\phi$ is the inflaton scalar field driving inflation
\cite{Mukh,Hawk}. If the field stays in the false vacuum state and does not
move, one has indefinitely large density perturbations, which makes the
corresponding part of inflation almost useless. Fortunately, in the new
inflation scenario the field $\phi$ does move during inflation, which under
certain conditions makes   density perturbations produced during inflation
relatively small.

However, new inflation was plagued by its own problems. Effective potential
with a flat plato near the origin is somewhat  artificial. In most  versions of
the new inflation scenario the inflaton field originally could not be in a
thermal equilibrium with other matter fields. The theory of high temperature
symmetry restoration, which was the basis for old and new inflation, simply did
not work in such a situation. Moreover,   thermal equilibrium  requires many
particles interacting with each other. This means that new inflation could
possibly explain why our universe was so large only if it was very large and
contained many particles from the very beginning. Finally, inflation in this
theory begins very late, and during the preceding epoch the universe could
easily collapse or become so inhomogeneous that inflation may never happen.
These problems have been well understood already in 1982. Unfortunately, there
is a certain inertia in the development of scientific theories, and even after
new inflation died there were many people who insisted that it is alive, and
there were many others who did not know that it is dead and tried to criticize
it. In particular, most of the critical remarks made about inflation by Bill
Unruh at this conference are related to this theory.

All these problems   were resolved with the introduction of the chaotic
inflation scenario \cite{Chaot}. In this scenario inflation may occur in the
theories with simplest potentials such as
$\pm m^2\phi^2 +\lambda \phi^4$. It may begin even if there was no thermal
equilibrium in the early universe, and it may start even at the Planckian
density, in which case the problem of initial conditions for inflation can be
easily resolved \cite{Chaot,MyBook}. The main  idea of chaotic inflation is
very simple and general. One should study all possible initial conditions
without assuming that the Universe was in a state of thermal equilibrium, and
that the field $\phi$ was in the minimum of its effective potential from the
very beginning. This scenario strongly deviated from the standard lore of the
hot big bang theory and was
psychologically difficult to accept.  Gradually, however, it became clear that
the idea of chaotic initial conditions is most general, and that it is much
easier to construct a consistent cosmological theory without insisting that
the existence of  thermal equilibrium and high temperature phase transitions in
the early universe is a necessary condition for inflation to occur.

\section{From the Big Bang theory to the theory of eternal inflation}

The next   step in the development of inflationary theory which I would like to
mention here is the discovery of the process of self-reproduction of
inflationary domains. This process was known to exist in old inflationary
theory \cite{Guth} and in the new one \cite{Vil}, but it is especially
surprising and leads to most profound consequences in the context of the
chaotic inflation scenario \cite{Eternal}. It appears that in many models large
scalar field during inflation produces large quantum fluctuations which may
locally increase the value of the scalar field in some parts of the universe.
These regions expand at a greater rate than their parent domains, and quantum
fluctuations inside them lead to production of new inflationary domains which
expand even faster. This surprising behavior leads to an eternal process of the
universe self-reproduction.

Thus during the last ten years inflationary theory changed considerably. It has
broken an umbilical cord connecting it with the old big
bang theory, and acquired an independent life of its own. For the
practical purposes of describing   the observable part of our Universe one may
still speak about the big bang, just as one can still use Newtonian gravity
theory to describe the Solar system with very high precision. However, if
one tries to understand the beginning of the Universe, or its end, or its
global structure, then some of the notions of the big bang theory become
inadequate. One of the main principles of the big bang theory is the
homogeneity of the Universe. The assertion of homogeneity seemed to
be so important that it was called  ``the cosmological principle.'' Without
using this principle it is hard to prove that the whole Universe appeared  {\it
at a single moment of time}, which was associated with the big bang. So far,
inflation remains the only theory which explains why the observable part of the
Universe is almost homogeneous. However, many versions of inflationary
cosmology predict that on a much larger scale the Universe should be extremely
inhomogeneous, with energy density varying from the Planck density to almost
zero. This is a consequence of the self-reproduction of the universe which we
just discussed. Instead of one single big bang producing a single-bubble
universe, we are speaking now about inflationary bubbles producing new bubbles,
producing new bubbles, {\it ad infinitum}. In the new theory there is no end of
the universe evolution, and the notion of the big bang    loses its dominant
position, being removed to the indefinite past.

{}From this new perspective many old problems of cosmology, including the
problem of initial conditions, look much less profound than they seemed before.
In many versions of inflationary theory it can be shown that the fraction of
the volume of the universe with given properties (with  given values of fields,
with a given density of matter, etc.)  does not depend on time,  both at the
stage of inflation and even after it. Thus each part of the universe evolves in
time, but the universe as a whole may be stationary, and the properties of its
parts do not depend on the initial conditions \cite{LLM}.

Of course, this happens only for the (rather broad) set of initial conditions
which lead to self-reproduction of the universe. However, only finite number of
observers live in the  universes  created in a state with initial conditions
which do not allow self-reproduction, whereas infinitely many observers live in
the  universes  with the conditions which allow self-reproduction. Thus it
seems plausible that we (if we are typical, and live in the place where most
observers do) should live in the universe  created in a state with initial
conditions which allow  self-reproduction. Incidentally, such initial
conditions appear with a much greater probability if one uses the tunneling
wave function of the universe \cite{Tunn} rather that the Hartle-Hawking one
\cite{HH}. On the other hand,  stationarity of the self-reproducing universe
implies that an exact knowledge of these initial conditions in a
self-reproducing universe is   irrelevant for the investigation of its future
evolution. In this sense the debate about the choice of the wave function
describing initial conditions looses its importance. One may   argue that even
the models of chaotic inflation with the potentials which do not allow
inflation near the Planck density (like the potentials used in new inflation)
become acceptable as far as they can support self-reproduction of the universe
\cite{LLM}.

During the process of self-reproduction, the universe becomes divided into
exponentially large domains containing matter in all possible ``phases"
corresponding to all possible vacuum states of the theory. Investigation of the
distribution of volume of different domains in the universe may give us a
possibility to explain geometric properties of space (including its
dimensionality and the type of compactification) and to find ``most probable"
values of coupling constants in our part of the universe
\cite{LLM,Pred,Center}. This possibility, however,   depends on the as-yet
unsolved problem of measure in
quantum cosmology.

\section {Inflation with $\Omega \not = 1$}

The new cosmological paradigm may have important implications not only for our
understanding of the global structure and the fate of inflationary universe,
but for observational cosmology as well.  For example, until very recently it
was believed that the universe  after inflation must become extremely flat,
with $\Omega = 1 \pm 10^{-4}$. If observational data will show that $\Omega$
differs from $1$ by more than a fraction of a percent, most of inflationary
models will be disproved.

Fortunately, it is possible to solve this problem, both for a closed universe
  and for an open one. The
main idea is to use the well known fact that the region of space created in the
process of a quantum tunneling tends to have a spherically symmetric shape,
and homogeneous interior, if the tunneling probability is suppressed strongly
enough. Then such bubbles of a new phase  tend  to evolve (expand) in a
spherically symmetric
fashion. Thus, if one
could associate the whole visible part of the universe with an interior of one
such region, one would solve the homogeneity problem, and then all other
problems
will be solved by the subsequent relatively short stage of inflation.

For a closed universe the realization of this program is relatively
straightforward \cite{Open}. One should consider the process of quantum
creation of  a
closed inflationary universe from ``nothing.''  If the probability of such a
process is exponentially suppressed (and this is indeed the case if inflation
is possible only at the energy density much smaller than the Planck density
\cite{Creation}), then the universe created that way will be  rather
homogeneous from the very beginning.

The situation with an open universe is much more complicated. Indeed, an open
universe is infinite, and it may seem impossible to create an infinite universe
by a tunneling process. However, this is not the case: any bubble formed in
the process of the false vacuum decay looks from inside like an infinite open
universe \cite{CL}.
 If this universe continues inflating
inside the bubble  then we obtain an open inflationary
universe.

There is an extensive investigation of the one-bubble open universe scenario.
However, until very recently it was not quite clear whether it is possible to
realize this scenario in a natural way.   An important step in this direction
was made when the first semi-realistic   models of  open inflation  were
proposed \cite{Turok}. These models were based on chaotic inflation and
tunneling in the theories of one scalar field $\phi$. However, as was shown in
\cite{Open}, in the natural versions of  such theories the tunneling occurs not
by bubble formation, but by jumping onto the top of the potential barrier
described by the Hawking-Moss instanton. This leads to formation of
inhomogeneous domains of a new phase, and the whole scenario fails. In order to
resolve this problem one is forced either to introduce very complicated
effective potentials, or consider
theories with nonminimal kinetic terms for the inflaton field.
This makes the models   fine-tuned and complicated. It  was   very
tempting to find a
more natural realization of the   inflationary universe scenario which would
give
inflation with $\Omega < 1$.

Fortunately, this goal
can be easily achieved if one considers models of two
scalar fields \cite{Open}. One of these fields may be the standard inflaton
field
$\phi$ with a relatively small mass, another may be, e.g., the scalar field
responsible for the symmetry breaking in GUTs. The presence of two scalar
fields allows one to obtain the required bending of the inflaton potential by
simply changing the definition of the inflaton field in the process of
inflation. At the first stage the role of the inflaton is played by a heavy
field with a steep barrier in its potential, while on the second stage the
role of the inflaton is played by a light field, rolling in a flat direction
``orthogonal'' to the direction of quantum tunneling. Inflationary models of
this type
are quite simple, yet they have many interesting features. In these models
the universe consists of infinitely many expanding bubbles immersed into
exponentially expanding false vacuum state. Each of these bubbles inside looks
like an infinitely large open universe, but the values of $\Omega$ in these
universes may take
any value from $1$ to $0$.  Thus we are again describing a self-reproducing
universe consisting of many (infinitely large)  universes with different
properties.

Here we will describe an extremely  simple model of two
scalar fields, where the universe after inflation becomes open  in a very
natural way. Consider a model of
two noninteracting scalar fields, $\phi$ and $\sigma$, with the effective
potential $V(\phi, \sigma) = {m^2\over 2}\phi^2 + V(\sigma)$.
Here $\phi$ is a weakly interacting inflaton field, and $\sigma$, for example,
can be the field responsible for the symmetry breaking in GUTs. We will assume
that $V(\sigma)$ has a local minimum at $\sigma = 0$, and a global minimum at
$\sigma_0 \not = 0$, just as in the old inflationary
theory. For definiteness, one may assume that this potential is given by
${M^2\over 2} \sigma^2 -
{\alpha M } \sigma^3 + {\lambda\over 4}\sigma^4 + V(0)$, with $V(0) \sim
{M^4\over 4 \lambda}$, but it is not essential;
no fine tuning of the shape of this potential is required.
Inflation begins at $V(\phi, \sigma) \sim M_{\rm P}^4$. At this stage
fluctuations of
both fields are very strong, and the universe enters the stage of
self-reproduction, which finishes for the field $\phi$ only when it becomes
smaller than $M_{\rm P} \sqrt{M_{\rm P}\over m}$ and the energy density drops
down to $m
M_{\rm P}^3$. Quantum fluctuations of
the field
$\sigma$ in some parts of the universe put it directly to the absolute minimum
of
$V(\sigma)$, but in some other parts the scalar field $\sigma$ appears in the
local minimum of $V(\sigma)$ at $\sigma  = 0$.

The main idea of our scenario can be explained as follows. Because the fields
$\sigma$ and
$\phi$ do not interact with each other, tunneling to the minimum of $V(\sigma)$
in different parts of the universe may occur   at different values of the field
$\phi$. The
parameters of the bubbles of the field $\sigma$ are determined by the mass
scale $M$ corresponding to the effective potential $V(\sigma)$. This mass scale
in our model is much greater than $m$. Thus the duration of tunneling in the
Euclidean ``time'' is much smaller than $m^{-1}$. Therefore the field $\phi$
practically does not change its value during the tunneling.  If
the probability of decay at a given $\phi$ is small enough, then it does not
destroy the whole vacuum state $\sigma = 0$; the bubbles of the new
phase are produced all the way when    the field $\phi$ rolls down to $\phi =
0$. In this process  the universe  becomes filled with
(nonoverlapping) bubbles immersed in the false vacuum state with $\sigma = 0$.
Interior of each of these bubbles   represents an open universe. However, these
bubbles   contain  {\it  different} values of the field $\phi$, depending on
the
value of this field at the  moment when the bubble formation occurred. If the
field $\phi$ inside a bubble is smaller than $3 M_{\rm P}$, then the universe
inside
this bubble will have a vanishingly small $\Omega$, at the age $10^{10}$ years
after the end of inflation it will be practically empty, and life of our type
would not exist there.  If the field $\phi$ is much greater than $3 M_{\rm P}$,
the
universe inside the bubble will be almost exactly flat, $\Omega = 1$, as in the
simplest version of the chaotic inflation scenario. It is important, however,
that  {\it  in  an eternally existing self-reproducing universe there will be
infinitely many universes containing any particular value of $\Omega$, from
$\Omega = 0$ to $\Omega = 1$}, and one does not need any fine tuning of the
effective potential to obtain a universe with, say,  $0.2 <\Omega <
0.3$.\footnote{The theories of two scalar fields  which allow a definite
prediction for $\Omega$ are also possible \cite{LidOc}.}

Should we take these models seriously? Should we admit that the standard
prediction of inflationary theory that $\Omega = 1$ is not universally valid? I
think that now it
is too late to discuss this question: the genie is already out of the bottle.
We know that inflationary models describing homogeneous inflationary universes
with $\Omega \not = 1$ do exist, whether we like it or not. It is still true
that the models which lead to
$\Omega = 1$ are much more abundant and, arguably, more natural.  However, in
our opinion, it is
very encouraging that
inflationary  theory is  versatile enough to include models with all
possible values of $\Omega$.

This situation makes some observers unhappy. Few years ago it seemed that they
can easily kill inflationary theory if they find  that the density of the
universe is not equal to the critical density. It would be a significant
scientific result.
Now the situation changed. If we find that $\Omega = 1$, it will be a
confirmation of inflationary theory because 99\% of inflationary models do
predict that $\Omega = 1$, and no other theories make this prediction. On the
other hand, if observations will show that $\Omega \not = 1$, it will not
disprove inflation.

This does not make inflationary theory untestable. Indeed,  each particular
inflationary model can be tested and ruled out by comparison of its predictions
with observational data.  In particular, the simplest model of open universe
which I just described needs to be modified: in its original form it predicts
too large anisotropy of the microwave background radiation \cite{Open,Sas}.
Fortunately, this problem disappears after a minor modification of the model
\cite{Open}:
$V(\phi,\sigma) = {g^2\over 2}\phi^2\sigma^2 + V(\sigma)$.
Thus, comparison with observations may rule out many versions of inflationary
theory which otherwise look rather attractive. However, it is very difficult
to disprove
 the basic idea of inflation.  But it is not our goal, is it?

To make my position more clear, I would like to discuss   the history of
the   standard model of electroweak interactions  \cite{GWS}.  Even though this
model was developed by  Glashow, Weinberg and Salam in the 60's, it became
popular only in 1972, when it was realized  that gauge theories with
spontaneous symmetry breaking are renormalizable \cite{Hooft}. However,   it
was immediately pointed out that this model is far from being perfect. In
particular, it  was not based on the simple group of symmetries, and it had
anomalies. Anomalies could destroy the renormalizability, and therefore it was
necessary to invoke a mechanism of their cancellation by enlarging the fermion
sector of the theory. This did not look very natural, and therefore Georgi and
Glashow in 1972 suggested another model \cite{GG}, which at the first glance
looked much better. It was based on the simple group of symmetry $O(3)$, and it
did not have any anomalies. In the beginning it seemed that this model is a
sure winner. However, after the discovery of neutral currents which could not
be described by the  Georgi-Glashow model, everybody forgot about the issues of
naturalness and simplicity and returned back to the  more complicated
Glashow-Weinberg-Salam model, which gradually became the standard model of
electroweak interactions.
This model has about twenty free parameters which so far did not find an
adequate theoretical explanation. Some of these parameters may  appear rather
unnatural. The best example is the coupling constant of the electron to the
Higgs field, which is $2\times 10^{-6}$. It is a pretty unnatural number which
is fine-tuned in such a way as to make the electron 2000 lighter than the
proton.
It is important, however, that all existing versions of the electroweak theory
are based on two fundamental principles: gauge invariance and spontaneous
symmetry breaking.   As
far as these principles hold, we can adjust our parameters and wait until they
get their interpretation in a context of a more general theory. This is the
standard way of development of the elementary particle physics.

For a long time cosmology developed in a somewhat different way, because of the
scarcity of reliable observational data.  Ten
years ago many different cosmological models (HDM, CDM, $\Omega = 1$, $\Omega
\ll 1$, etc.) could describe all observational data reasonably well. The main
criterion for a good theory was its beauty and naturalness. Now it
becomes increasingly complicated to explain all observational data. Therefore
cosmology is gradually becoming a normal experimental science, where the
results of observations play a more important role  than the considerations of
naturalness. However, in our search  for a correct   theory we cannot give up
the requirement of its internal consistency. In particle physics the main
principle  which made this theory internally consistent was gauge
invariance.  It seems that in cosmology
something like inflation is needed to make the universe large and homogeneous.
It is true that most of the  inflationary models
predict a  universe with $\Omega  = 1$.  Hopefully,  several years later we
will know that our universe is flat, which will be a strong experimental
evidence in favor of  inflationary cosmology in its simplest form. However, if
observational data will show,
beyond any reasonable doubt, that $\Omega \not = 1$, it will not imply  that
inflationary theory is wrong, just like the discovery of neutral currents did
not disprove gauge theories of electroweak interactions.  Indeed, the only
consistent theory of a {\it large homogeneous} universe with $\Omega \not = 1$
which is available now is based on inflationary cosmology.

Thus, by measuring $\Omega$ we may rule out a large class of inflationary
models but we will be unable to rule out the idea of inflation.  What's about
microwave background anisotropy and the theory of large scale structure of the
universe? Again, it is possible to confirm inflationary models because most of
them predict perturbations with specific properties. However, if it so happens
that inflationary perturbations are in a conflict with observational data, then
one can easily propose inflationary models where inflation solves homogeneity,
isotropy and other problems, but produces extremely small density
perturbations. Then one will be free to use his best non-inflationary mechanism
for production of density perturbations (strings, textures, etc.).  They can be
created, e.g., by nonthermal phase transitions after explosive reheating
\cite{Nonthermal}.

Perhaps there is a  way to disprove inflationary cosmology by comparing its
predictions with observational data: If we   find that the universe rotates as
a whole, or has any other anisotropy which is described by vector perturbations
of metric, then it will be very difficult to make it compatible with inflation
\cite{Ellis}. Indeed we know that scalar and tensor perturbations can be
produced during inflation, but vector perturbations, just like other vector
fields, can hardly be produced. But what if we find out a nontrivial way of
producing vector perturbations, just like we found a way to produce an
inflationary universe with $\Omega \not = 1$?   It seems that the only sure way
to kill inflationary cosmology (if one really wants to do it) is to suggest a
better cosmological theory.

\end{document}